\documentclass[12pt, a4paper]{article}
\usepackage{epsf}
\usepackage{cite}
\usepackage{amsmath,amssymb}
\input{colordvi.tex}
\usepackage[usenames,dvipsnames]{color}
\usepackage{graphicx}
\usepackage{ascmac}
\usepackage{hyperref}
\usepackage{slashed}
\usepackage{tikz-feynman}

\begin{document}

\begin{titlepage}

\begin{center}
\hfill TU-1239\\
\hfill KEK-QUP-2024-0017
\vskip 1.in

\renewcommand{\thefootnote}{\fnsymbol{footnote}}

{\Large \bf
Revisiting the Minimal Nelson-Barr Model
}

\vskip .5in

{\large
Kai Murai$^{(a)}$\footnote{kai.murai.e2@tohoku.ac.jp}
and
Kazunori Nakayama$^{(a,b)}$\footnote{kazunori.nakayama.d3@tohoku.ac.jp}
}

\vskip 0.5in

$^{(a)}${\em 
Department of Physics, Tohoku University, Sendai 980-8578, Japan
}

\vskip 0.2in

$^{(b)}${\em 
International Center for Quantum-field Measurement Systems for Studies of the Universe and Particles (QUP), KEK, 1-1 Oho, Tsukuba, Ibaraki 305-0801, Japan
}

\end{center}
\vskip .5in

\begin{abstract}

We revisit the minimal Nelson-Barr model for solving the strong CP problem through the idea of spontaneous CP breaking. The minimal model suffers from the quality problem, which means that the strong CP angle is generated by higher-dimensional operators and one-loop effects. 
Consequently, it has been considered that there is a cosmological domain wall problem and that leptogenesis does not work. We point out that just imposing an additional approximate global symmetry solves the quality problem. We also propose a simple solution to the domain wall problem and show that the thermal leptogenesis scenario works.

\end{abstract}

\end{titlepage}

\tableofcontents

\renewcommand{\thefootnote}{\arabic{footnote}}
\setcounter{footnote}{0}

\section{Introduction}

The strong CP problem is one of the most puzzling issues remaining in the Standard Model (see, e.g., Refs.~\cite{Kim:2008hd,Hook:2018dlk} for reviews).
Theoretically, quantum chromodynamics (QCD) includes a CP-violating term $\propto \theta_s \mathrm{Tr}[G_{\mu\nu} \tilde{G}^{\mu\nu}]$ with $G_{\mu\nu}$ and $\tilde{G}^{\mu\nu}$ being the field strength of the gluon and its dual, respectively.
Taking into account the chiral rotation of the quark phases, the CP violation is parameterized by the invariant angle:
\begin{align}
    \bar{\theta}_s 
    \equiv
    \theta_s + \arg [\det ({\textbf m}_u {\textbf m}_d)]
    \ ,
\end{align}
where ${\textbf m}_u$ and ${\textbf m}_d$ are the mass matrices of the up and down-type quarks, respectively.
Since $\theta_s$ is a free parameter of the Standard Model, $\bar{\theta}_s$ is expected to be of $\mathcal{O}(1)$ from naturalness.
On the other hand, the measurement of the electric dipole moment of the neutron gives a stringent limit, $|\bar{\theta}_s| \lesssim 10^{-10}$~\cite{Abel:2020pzs}.

As a solution to the strong CP problem, two classes of models are well-known:%
\footnote{There is also a solution with the massless up-quark~\cite{Georgi:1981be,Kaplan:1986ru,Choi:1988sy,Banks:1994yg}, which is currently disfavored by lattice QCD~\cite{Fodor:2016bgu,Alexandrou:2020bkd,FlavourLatticeAveragingGroupFLAG:2021npn}.}
the Peccei-Quinn mechanism~\cite{Peccei:1977hh,Peccei:1977ur} accompanied by the axion~\cite{Weinberg:1977ma,Wilczek:1977pj} and the spontaneous or soft breaking of a discrete spacetime symmetry such as parity~\cite{Beg:1978mt,Mohapatra:1978fy,Babu:1989rb,Barr:1991qx} or CP symmetry~\cite{Georgi:1978xz,Segre:1979dx,Barr:1979as,Nelson:1983zb,Barr:1984qx,Bento:1991ez}.
In this paper, we focus on the latter class, in particular, the Nelson-Barr mechanism~\cite{Nelson:1983zb,Barr:1984qx}.

In the Nelson-Barr mechanism, CP is an exact symmetry of the fundamental theory, and thus $\bar{\theta}_s = 0$ at the Lagrangian level.
To reproduce the CP-violating angle in the Cabibbo-Kobayashi-Maskawa (CKM) matrix, CP symmetry must be spontaneously broken by certain fields. 
Although this breaking induces complex phases in the quark mass matrix, the smallness of $\bar{\theta}_s$ is preserved due to a specific structure of the mass matrix.
Despite being proposed several decades ago, the class of the Nelson-Barr model and its extensions remain an active field of research even in this decades (see, e.g., Refs.~\cite{Vecchi:2014hpa,Dine:2015jga,Davidi:2017gir,Schwichtenberg:2018aqc,Cherchiglia:2019gll,Evans:2020vil,Cherchiglia:2020kut,Perez:2020dbw,Cherchiglia:2021vhe,Valenti:2021rdu,Valenti:2021xjp,Fujikura:2022sot,Girmohanta:2022giy,McNamara:2022lrw,Asadi:2022vys,Perez:2023zin,Suematsu:2023jqa,Banno:2023yrd,Dine:2024bxv,Bastos:2024afz}).

The simplest viable model for the Nelson-Barr mechanism was proposed by Bento, Branco, and Parada (BBP)~\cite{Bento:1991ez}.
The BBP model includes a complex scalar and vector-like quarks, which are odd under a $Z_2$ symmetry.
In the BBP model and its extensions, the scale of spontaneous CP breaking should be lower than about $\sim 10^8$\,GeV in order to avoid the too large contribution to the strong CP angle $\theta_s$ due to the higher dimensional operator (see Eq.~(\ref{higherdim})). 
This may make the BBP model incompatible with the leptogenesis scenario~\cite{Fukugita:1986hr} since the leptogenesis demands the temperature of the universe higher than $\sim 10^9$\,GeV~\cite{Giudice:2003jh,Buchmuller:2004nz} while such a high temperature would lead to the domain wall formation in association with the spontaneous CP breaking.
To successfully implement leptogenesis in the framework of the BBP model, several studies~\cite{Asadi:2022vys,Suematsu:2023jqa} extend both the symmetry and field contents.

In this paper, we consider the same setup as the minimal BBP model in terms of the field content.
By imposing an additional approximate discrete global symmetry, some mass and interaction terms in the original BBP model are suppressed by a small parameter.
Then, we can realize a high scale of spontaneous CP breaking $\gg 10^8$\,GeV evading the quality problem of the Nelson-Barr mechanism while maintaining the $\mathcal{O}(1)$ CKM angle.
Even if the reheating temperature is higher than the CP breaking scale, CP symmetry can be spontaneously broken during and after inflation by introducing negative Hubble and thermal potentials for the CP-violating scalar.
Consequently, leptogenesis can be successfully realized since the CP violation is also introduced to the lepton sector by assigning appropriate charges under the discrete symmetries to the Standard Model leptons and right-handed neutrinos.

The rest of this paper is organized as follows.
In Sec.~\ref{sec:minimal}, we briefly review the minimal Nelson-Barr model, i.e., the BBP model, and introduce an additional symmetry.
In Sec.~\ref{sec:cos}, we discuss the formation of domain walls and leptogenesis in our model.
Finally, we conclude and discuss our results in Sec.~\ref{sec:conc}.

\section{Minimal Nelson-Barr model}
 \label{sec:minimal}

\subsection{BBP model}

A ``minimal'' model of spontaneous CP violation was proposed in Ref.~\cite{Bento:1991ez}, which we call the BBP model.
In the BBP model, we introduce vector-like quarks
$D_L$ and $D_R$ with the U(1) hypercharge $-1/3$ and a singlet complex scalar $S$ and impose CP symmetry.
Furthermore, a $Z_2$ symmetry is imposed under which $D_L, D_R$, and $S$ are odd while all other fields are even.
The down-quark sector and the scalar sector of the model are given by
\begin{align}
	\mathcal L =-\left[ M\overline D_L D_R +(g_i S + g_i' S^*) \overline D_L d_{Ri} + y^d_{ij} H \overline Q_{Li} d_{Rj} + {\rm h.c.}\right] - V(S,H),
	\label{BBP}
\end{align}
where $M$, $g_i$, $g_i'$, $y_{ij}^d$ $(i,j=1,2,3)$ are real constants due to the CP symmetry, and $d_{Ri}, Q_{L_i}, H$ are the Standard Model right-handed down-type quarks, left-handed quark doublets, and Higgs doublet.
Note that the term like $H \overline Q_{Li} D_R$ is forbidden by the $Z_2$ symmetry.
The scalar potential is given by
\begin{align}
	V(S,H) = &\lambda_H\left(|H|^2-v_H^2\right)^2 +  \lambda_S\left(|S|^2-v_S^2\right)^2 - \lambda_{SH}\left(|H|^2-v_H^2\right)\left(|S|^2-v_S^2\right) \nonumber \\ 
	&+ (\mu^2 + \gamma_{SH} |H|^2 + \gamma_{2}|S|^2)(S^2+S^{*2}) + \gamma_4 (S^4+ S^{*4}),
 \label{VSH}
\end{align}
where all coefficients are real. The second line gives the potential for the angular component of $S$.
For the moment, just for simplicity, we assume that the second line is small compared to the first line so that the vacuum expectation values (VEVs) of $|H|$ and $|S|$ are mostly determined by the first line.
We require $\lambda_H>0$ and $4\lambda_H \lambda_S > \lambda_{SH}^2$ so that $|H| = v_H$ and $|S|=v_S$ is the minimum of the potential.
Later we will see that $\lambda_{SH} > 0$ is favored from the cosmological viewpoint.
With this assumption, the VEV of $S$ is given by $S = v_S e^{i\theta}$ with $\theta$ being an arbitrary value depending on the choice of $\gamma_2$ and $\gamma_4$. 

The mass matrix of the quarks is given by
\begin{align}
	-\mathcal L = (\overline d_{Li}, \overline D_L) \mathcal M 
	\begin{pmatrix}
		d_{R j} \\ D_R	
	\end{pmatrix} + {\rm h.c.},~~~~~~
	\mathcal M = \begin{pmatrix}
		{\textbf m}_{ij} &  0 \\ B_j  & M	
	\end{pmatrix},
	\label{massmatrix}
\end{align}
where ${\textbf m}_{ij} \equiv y^d_{ij} v_H$ and $B_j \equiv (g_j e^{i\theta} + g_j' e^{-i\theta} ) v_S$. It is only $B_i$ that is complex in the mass matrix. 
However, $B_i$ does not contribute to the determinant of the mass matrix, hence ${\rm arg}[{\rm det}\mathcal M ]= 0$.
Thus there is no strong CP angle at the tree level.
On the other hand, the CP phase in the CKM matrix appears after the diagonalization of this mass matrix.

Let us see how the consistent CKM matrix is obtained in this model.
The mass matrix (\ref{massmatrix}) is diagonalized by the bi-unitary transformation as $d_{R\alpha} = U_{\alpha\beta} d'_{R\beta}$ and $d_{L\alpha} = V_{\alpha\beta} d'_{L\beta}$, where the prime denotes mass eigenstates and $\alpha, \beta =1$--$4$.
Here, $\alpha=1$--$3$ corresponds to the Standard Model down quarks, and $d_4 = D$ is the heavy quark.
The mass matrix squared is diagonalized as
\begin{align}
	V^\dagger \mathcal M \mathcal M^\dagger V = {\rm diag}(m_d^2, m_s^2, m_b^2, m_D^2),
\end{align}
where $m_d,m_s, m_b, m_D$ are down, strange, bottom, and additional heavy quark masses, respectively. By explicitly writing the unitary matrix as
\begin{align}
	V = \begin{pmatrix} (V_1)_{ij} & (V_{2})_i \\ (V_3)_j & V_4  \end{pmatrix},
\end{align}
and assuming $m_D^2, M^2, |B_i|^2 \gg m_i^2, {\textbf m^2}$, we obtain
\begin{align}
	m_D^2 \simeq |B|^2 + M^2,~~~~~~V_2 \simeq \frac{{\textbf m} B^\dagger}{m_D^2}V_4,~~~~~~V_3 \simeq -\frac{B {\textbf m}^{\rm T}}{m_D^2} V_1,	
\end{align}
where $|B|^2=BB^\dagger = \sum_i |B_i|^2$.
The $3\times 3$ submatrix $V_1$, combined with the unitary matrix from the up-quark sector, defines the CKM matrix for the weak interaction process. $V_1$ is not exactly unitary, but its deviation from the unitarity is suppressed as $V_1^\dagger V_1 \simeq {\textbf I} + \mathcal O(|B{\textbf m}^\mathrm{T}|^2 / m_D^4)$. 
We also find 
\begin{align}
	V_1\,{\rm diag}(m_d^2, m_s^2, m_b^2)\,V_1^\dagger \simeq {\textbf m} {\textbf m}^{\rm T} - \frac{{\textbf m}B^\dagger B {\textbf m}^{\rm T}}{m_D^2}.
\end{align}
The first and second terms on the right-hand side are of the same order if $m_D^2 \sim |B|^2$. Note that the second term includes complex phases since $B_i$ is in general complex. This indicates that the matrix $V_1$ also contains $\mathcal O(1)$ complex phases if $m_D^2 \sim |B|^2$.  
Thus below we assume $|M| \sim |B_i|$ to obtain $\mathcal O(1)$ CP phase in the CKM matrix.\footnote{
	One of the three phases of $B_i$ $(i=1-3)$ can be removed by the phase redefinition of $D_L$. Thus at least two components $B_i$ should have different phases.
}

There is a ``quality problem'' in this model. 
In the original BBP model~\cite{Bento:1991ez}, it is allowed to introduce the higher-dimensional operator
\begin{align}
	\mathcal L = \frac{c_iS + c_i'S^*}{\Lambda} H \overline Q_{Li} D_R + {\rm h.c.},
	\label{higherdim}
\end{align}
with cutoff scale $\Lambda$ and $c_i, c_i'$ being real constants. This term contributes to the upper right entry of the mass matrix $\mathcal M$ (\ref{massmatrix}), and hence the strong CP angle is induced.
The strong CP angle is estimated as $\bar\theta_s \sim v_S/\Lambda$, which should be smaller than $\sim 10^{-10}$.
If $\Lambda \sim M_{\rm Pl}$, this requires $v_S \lesssim 10^8\,$GeV.
Due to this constraint, the BBP model suffers from several cosmological difficulties. 
One is the cosmological domain wall problem in association with the spontaneous CP breaking~\cite{McNamara:2022lrw}, since the symmetry may be likely restored in the early universe. 
If the maximum cosmic temperature is sufficiently low, then the domain walls are inflated away during inflation and there is no domain wall problem, but then it is difficult to explain the baryon asymmetry of the universe.

Furthermore, the strong CP angle is induced at the loop level. 
The quark mass matrix receives quantum correction that depends on the VEV of $S$, which in turn gives the strong CP angle. The detailed calculation is found in App.~\ref{app:oneloop}.
Its typical order is estimated as
\begin{align}
	\bar\theta_s \sim \frac{1}{32\pi^2} \gamma_{SH}\sin(2\theta)\sum_i(g_i^2 -g_i'^2) \log\left(\frac{m_h^2}{m_\sigma^2}\right),
\end{align}
where $m_h$ and $m_\sigma$ are the masses of the Higgs particle and the angular component of $S$, respectively.
Thus either $\gamma_{SH}$ or $g_i, g_i'$ must be sufficiently small to avoid the regeneration of the strong CP angle.

\subsection{BBP model with additional symmetry}

Now let us slightly modify the BBP model. 
The Lagrangian is the same as (\ref{BBP}), but now the charge assignments are modified as shown in Table~\ref{table}. 
We impose a $Z_4$ symmetry, instead of the $Z_2$ symmetry in the original proposal, so that the model should be consistent with the leptogenesis, as will be shown in the next section.
In addition, we impose an approximate $Z_{4n}$ symmetry with $n$ an integer.
The terms involving $\overline D_L$ in (\ref{BBP}) are actually inconsistent with this $Z_{4n}^{\rm (app)}$ symmetry.
Thus the parameters $M$, $g_i$, and $g_i'$ should be understood as the small symmetry-breaking parameter.
To make this point clearer, we introduce a parameter $\epsilon \ll 1$ that characterizes the small explicit breaking of $Z_{4n}^{\rm (app)}$ symmetry:
\begin{align}
	\mathcal L =-\left[ \epsilon^k M\overline D_L D_R +\epsilon^k (g_i S + g_i' S^*) \overline D_L d_{Ri} + y^d_{ij} H \overline Q_{Li} d_{Rj} + {\rm h.c.}\right] - V(S,H).
\end{align}
with $k$ being some integer in association with the quark charge assignments as given in Table~\ref{table}.
The parameter $\epsilon$ may be regarded as a spurion field that would have a $\pm 1$ charge under the $Z_{4n}^{\rm (app)}$ symmetry.
Note that all the terms in the scalar potential (\ref{VSH}) are consistent with $Z_{4n}^{\rm (app)}$, and hence all the parameters in the potential are not suppressed by $\epsilon$. 
With this parametrization, we assume $M \sim v_S$, $g_i \sim g_i' \sim \mathcal O(1)$.
Further, we assume $M \gg v_H$ so that $\epsilon^k M \gg v_H$ is satisfied.
Then the $\mathcal O(1)$ CP angle in the CKM matrix is retained.

With these charge assignments, the one-loop contribution to the strong CP angle is given by 
\begin{align}
	\bar\theta_s \sim \frac{\epsilon^{2k}}{32\pi^2} \gamma_{SH}\sin(2\theta)\sum_i(g_i^2 -g_i'^2) \log\left(\frac{m_h^2}{m_\sigma^2}\right).
\end{align}
Thus the strong CP angle is small enough for $\epsilon^k \lesssim 10^{-5}$.
In contrast to the original model, the suppression of the loop contribution can be understood as a consequence of the smallness of the symmetry-breaking parameter.
The higher-dimensional operator (\ref{higherdim}) should also be multiplied by $\epsilon^{2k}$ factor:
\begin{align}
	-\mathcal L = \epsilon^{2k}\frac{c_iS + c_i'S^*}{\Lambda} H \overline Q_{Li} D_R + {\rm h.c.},
    \label{eq: higher operator}
\end{align}
thus it is safely negligible for $\epsilon^k \lesssim 10^{-5}$ even if $v_S/\Lambda$ is close to unity.
In addition, higher-dimensional operators involving $S^2/\Lambda^2$ or $S^{* 2}/\Lambda^2$ are allowed by the $Z_4$ and $Z_{4n}^\mathrm{(app)}$ symmetries. Thus all the terms written so far multiplied by these factors are also allowed. These terms can induce complex corrections to the quark mass matrix with the contribution of order $(v_S/\Lambda)^2$.\footnote{A term proportional to $\epsilon^k S^2\overline D_L D_R$ should be understood as $(S/\Lambda)^2\times\epsilon^k M \overline D_L D_R$ since we already assumed a mass parameter $M$ that is much smaller than the cutoff scale.
Thus we require $ v_S/\Lambda \lesssim 10^{-5}$.}
Now we are allowed to have a large VEV $v_S \gg 10^8\,$GeV for sufficiently large $\Lambda$, in contrast to the original model.
As we will see in the next section, such a large $v_S$ makes the leptogenesis scenario viable while the cosmological domain wall problems in association with the spontaneous CP breaking can be easily avoided.
\begin{table}[t]
\centering
\begin{tabular}{|c|c|c|c|c|c|c|c|c|c|c|} \hline
~          & $S$ & $D_L$ & $D_R$ & $Q_{Li}$ & $d_{Ri}$ & $u_{Ri}$ & $H$ & $N_i$ & $L_i$ & $e_{Ri}$ \\ \hline
$Z_4$  & $2$ & $2$ & $2$ & $0$ & $0$ & $0$ & $0$ & $1$ & $1$ & $1$ \\ \hline
$Z^{\rm (app)}_{4n}$  & $2n$ & $2n$ & $2n-k$ 
& $k$ & $k$ & $k$ & $0$ & $n$ & $n$ & $n$\\ \hline
\end{tabular}
\caption{Charge assignments on the fields in the modified model. }
\label{table}
\end{table}

\begin{figure}[t]
    \centering
    \includegraphics[width=.7\textwidth ]{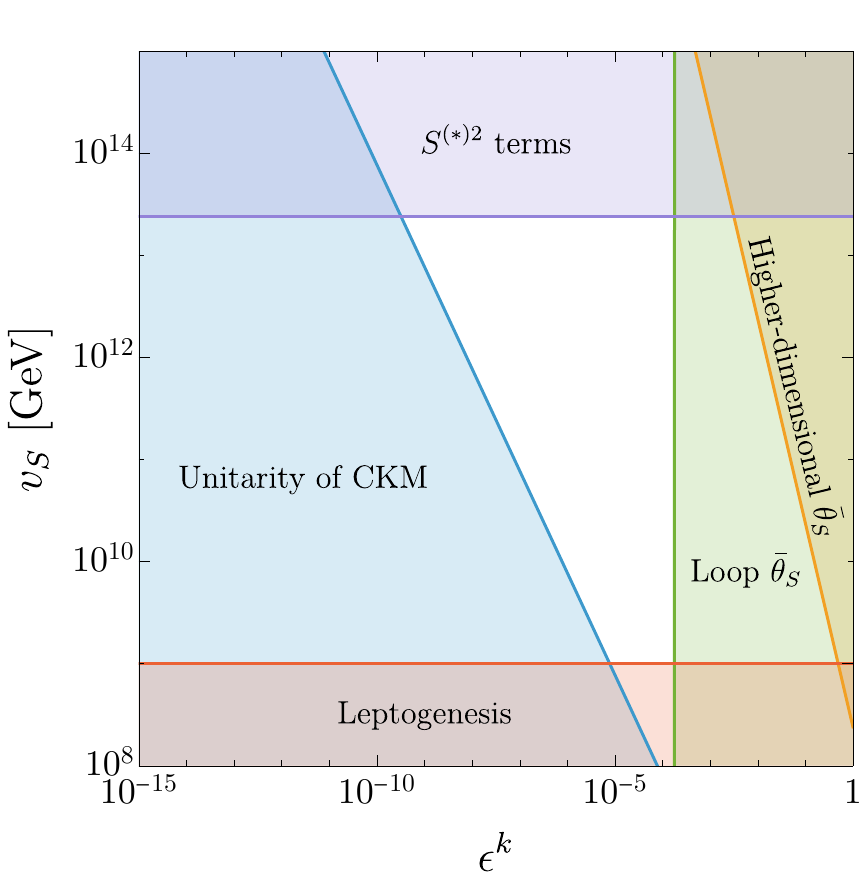}
    \caption{%
        Constraints and viable parameter region in the modified BBP model.
        For simplicity, we use $v_S = M$ and $\Lambda = M_\mathrm{Pl}$.
        The blue, orange, and green regions are constrained by the unitarity of the CKM matrix, $\bar{\theta}_s$ from the higher-dimensional operator~\eqref{eq: higher operator}, and $\bar{\theta}_s$ from the loop contribution, respectively.
        Above the violet line, higher-dimensional operators with $S^{(*)2}/\Lambda^2$ may induce $\bar{\theta}_s \gtrsim 10^{-10}$, and the red line comes from the requirement for leptogenesis as discussed later.
    }
    \label{fig: epsilon vs vS}
\end{figure}

We show the constraints and viable parameter region in this model in Fig.~\ref{fig: epsilon vs vS}.
Here, we use $v_S = M$ and $\Lambda = M_\mathrm{Pl}$ and neglect $\mathcal{O}(1)$ parameters for simplicity.
In particular, we require $V_1^\dagger V_1 - {\textbf I} \sim v_H^2/(\epsilon^{2k} v_S^2) \lesssim 10^{-3}$ from the unitarity of the CKM matrix~\cite{ParticleDataGroup:2024cfk}.
If the parameters are near the edge of this constraint, this model can be probed via the deviation from the unitarity.
From the smallness of $\bar{\theta}_s$, we require $\epsilon^{2k}/(32\pi^2) \lesssim 10^{-10}$ and $\epsilon^{2k}v_S/\Lambda \lesssim 10^{-10}$.
As we will discuss below, thermal leptogenesis can be successful for $v_S \gtrsim 10^9$\,GeV.
In addition, the mass of vector-like quarks is constrained as $\epsilon^k M \gtrsim 1$\,TeV from collider searches (see, e.g., Refs.~\cite{ParticleDataGroup:2024cfk,ATLAS:2024fdw,CMS:2024bni,Banerjee:2024zvg} and references therein), which gives a looser limit than that from the unitarity of the CKM matrix.
If we do not require successful leptogenesis and neglect $\bar{\theta}_s$ from the loop contribution, we find viable parameters at $\epsilon^k = 1$ and $v_S \lesssim 10^8$\,GeV, which correspond to the original BBP model.

\section{Cosmology of minimal Nelson-Barr model}
 \label{sec:cos}

\subsection{Avoiding domain wall formation}
\label{sec:DW}

In this section, we consider the cosmological dynamics of the scalar field $S$. If it is in a symmetric phase $S=0$ during inflation and the symmetry breaking happens after inflation, domain walls are formed, and it leads to a cosmological disaster.
However, below we will see that it is possible that the symmetry is always broken $(\left<S\right>\neq 0)$ throughout the history of the universe, and hence the model is free from the domain wall problem.

The potential of $S$ in the early universe is given by
\begin{align}
	V(S) \simeq -(2\lambda_S v_S^2 + c \mathcal H^2)|S|^2 - \lambda_{SH}|H|^2|S|^2 + \lambda_S |S|^4,
 \label{VS}
\end{align}
where $\mathcal H$ is the Hubble parameter, and $c$ is assumed to be a positive constant of order unity.\footnote{
    The negative Hubble mass term arises if there is coupling like $\mathcal L = c\frac{R}{12}|S|^2$ with $R$ being the Ricci scalar.
}
During inflation, the $S$ field is placed at $S_{\rm min} = \sqrt{\frac{c}{2\lambda_S}}\mathcal H$, while the Higgs field $H$ sits at the origin $H=0$ by assuming the positive Hubble mass term for the Higgs.
Thus the CP symmetry is already broken during inflation in this case and domain walls are inflated away.

The issue is whether the symmetry is restored or not after inflation ends.
Let us consider a standard scenario in which the inflaton begins a coherent oscillation after inflation ends and gradually decays to radiation, and finally the radiation-dominated universe begins at the temperature $T=T_{\rm R}$.
Since the Higgs field is thermalized, the coupling with the Higgs leads to a negative thermal mass term to $S$,
\begin{align}
	V(S) \simeq -(2\lambda_S v_S^2 + c \mathcal H^2)|S|^2 - \frac{\lambda_{SH}}{6}T^2|S|^2 + \lambda_S |S|^4.
 \label{VS thermal}
\end{align}
Here an important assumption is that $\lambda_{SH}$ is positive.
The minimum of the potential is $S_{\rm min} \simeq \sqrt{\frac{\lambda_{SH}}{12 \lambda_S}}T$ for $T \gtrsim \sqrt{\frac{12\lambda_{S}}{\lambda_{SH}}}v_S$.
Around this temperature-dependent temporal minimum, $S$ itself may be thermalized, and a positive contribution to the thermal mass arises from the $S$ self-coupling. Then we require $\lambda_{SH} > 2\lambda_S$ for the negative thermal mass to dominate.
Note that the Higgs field also obtains a negative mass term of the order of $T$, but still the Higgs field is placed at the origin due to the thermal mass contributed from other Standard Model particles.
Thus the symmetry breaking happens at high temperatures, which is sometimes called the inverse symmetry breaking~\cite{Weinberg:1974hy,Jansen:1998rj,Pinto:1999pg}.

To avoid the dangerous domain wall formation, we must ensure that the $S$ field does not overshoot the origin $S=0$ until $T$ drops well below $v_S$.
One way to check this is to compare the time dependence of the ratio $|\delta S| / S_{\rm min}$, where $|\delta S|$ denotes the amplitude of the fluctuation around the minimum $S_{\rm min}$. 
If it is a decreasing function of time, $S$ never overshoots the potential origin $S=0$.
By noting that $T(\delta S)^2 \propto a^{-3}$, we have
\begin{align}
	\frac{|\delta S|}{S_{\rm min}} \propto \begin{cases}
            a^{-15/16} & {\rm for}~T>T_{\rm R} \\
            {\rm const.} & {\rm for}~T<T_{\rm R} 
        \end{cases}.
        \label{eq: deltaS/Smin}
\end{align}
Thus actually it is decreasing before the completion of the reheating.\footnote{
    If there were no thermal mass term and the negative Hubble mass term determines the temporal minimum of the potential, this ratio does not decrease for $S^4$ potential~\cite{Dine:1995kz}, and hence it likely leads to domain wall formation (see e.g. Ref.~\cite{Ema:2015dza}).
}
Below we confirm this observation with numerical calculations.

We assume that the inflaton behaves as non-relativistic matter and decays into the Standard Model plasma by a constant decay rate $\Gamma_\phi$ during the reheating epoch.
Then, the evolution of the energy density is given by 
\begin{align}
    \dot{\rho}_\phi + 3 \mathcal{H} \rho_\phi 
    &= 
    -\Gamma_\phi \rho_\phi
    \nonumber \\
    \dot{\rho}_\mathrm{r} + 4 \mathcal{H} \rho_\mathrm{r} 
    &= 
    \Gamma_\phi \rho_\phi
    \ ,
\end{align}
where $\rho_\phi$ and $\rho_\mathrm{r}$ are the energy densities of the inflaton and radiation, respectively, and $\mathcal{H}$ satisfies
\begin{align}
    3 M_\mathrm{Pl}^2 \mathcal{H}^2
    =
    \rho_\phi + \rho_\mathrm{r}
    \ .
\end{align}
Just after inflation, the energy densities are given by $\rho_\phi \simeq 3 M_\mathrm{Pl}^2 \mathcal{H}_\mathrm{end}^2$ and $\rho_\mathrm{r} \simeq 0$ with $\mathcal{H}_\mathrm{end}$ being the Hubble parameter at the end of inflation.

During the reheating epoch, the Standard Model plasma is thermalized, and then its temperature $T$ is defined by 
\begin{align}
    \rho_\mathrm{r}
    =
    \frac{\pi^2 g_*}{30}T^4 
    \ ,
\end{align}
where $g_*$ is the relativistic degrees of freedom at high temperatures.
In the following, we adopt $g_* = 106.75$ as a typical value.
After inflation, $T$ rapidly increases and takes the maximum value, $T_\mathrm{max} \sim (M_\mathrm{Pl}^2 \mathcal{H}_\mathrm{end} \Gamma_\phi)^{1/4}$, and then decreases due to the cosmic expansion.
Finally, the reheating epoch terminates when the Hubble parameter satisfies
\begin{align}
    \Gamma_\phi 
    =
    3\mathcal{H}_\mathrm{R}
    \ .
\end{align}
Then, the reheating temperature, $T_\mathrm{R}$, is determined by
\begin{align}
    \frac{\pi^2 g_*}{30}T_\mathrm{R}^4 
    =
    \frac{M_\mathrm{Pl}^2 \Gamma_\phi^2}{3}
    \ .
\end{align}
From these relations, we obtain the time evolution of $T$ and $\mathcal{H}$ after inflation.
Using them, we solve the equation of motion for $s \equiv \sqrt{2}|S|$:
\begin{align}
    \ddot{s} + 3 H \dot{s} 
    - \left( \mu_S^2 + c\mathcal{H}^2 + \frac{\lambda_{SH}T^2}{6} \right) s + \lambda_S s^3
    = 0
    \ ,
\end{align}
where $\mu_S^2 \equiv 2 \lambda_S v_S^2$.
As an initial condition, we set $s = s_\mathrm{min,end}$ and $\dot{s} = 0$ at the end of inflation, where $s_\mathrm{min,end}$ is $s_\mathrm{min} \equiv \sqrt{2}S_\mathrm{min}$ at the end of inflation.

\begin{figure}[t]
    \centering
    \includegraphics[width=.45\textwidth ]{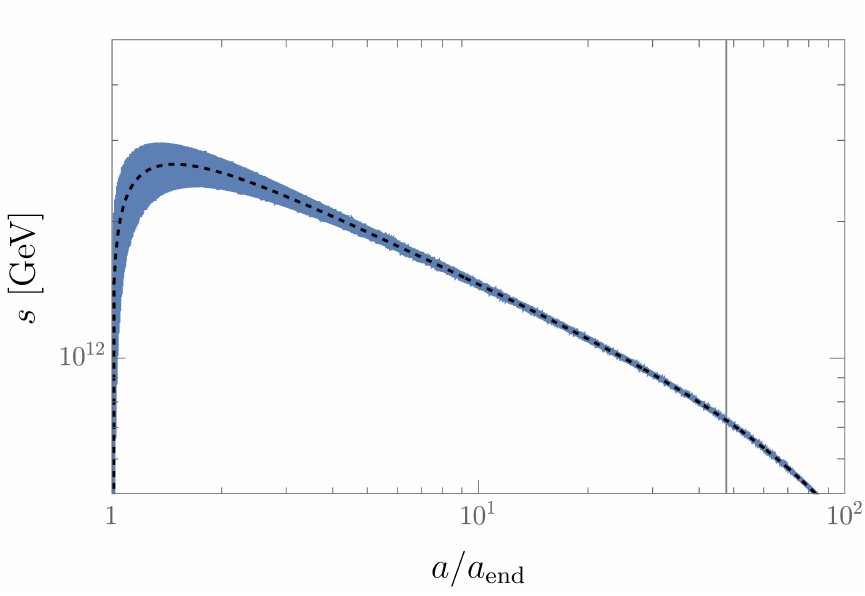}
    \hspace{5mm}
    \includegraphics[width=.45\textwidth ]{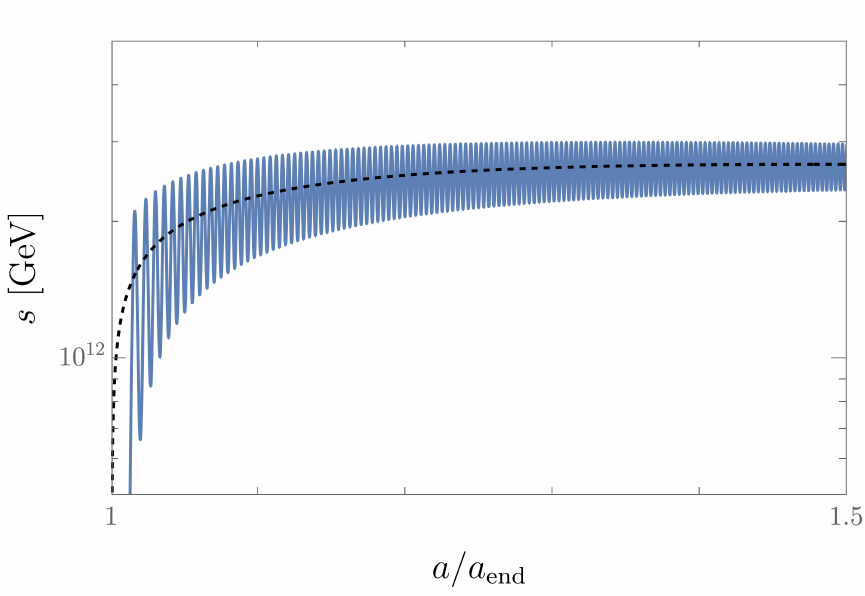}
    \caption{%
        Time evolution of $s$ for $\mathcal{H}_\mathrm{end} = 10^9$\,GeV, $T_\mathrm{R} = 10^{12}$\,GeV, $\mu_S = 10^5$\,GeV, $c = 1$, $\lambda_S = 0.2$, and $\lambda_{SH} = 0.5$.
        The horizontal axis is the scale factor normalized by that at the end of inflation, $a_\mathrm{end}$.
        The vertical gray line corresponds to $T = T_\mathrm{R}$.
        The black-dashed lines show $s_\mathrm{min}$.
        The right panel is an enlarged view of $1 \leq a/a_\mathrm{end} \leq 1.5$.
    }
    \label{fig: time evolution of s}
\end{figure}

\begin{figure}[t]
    \centering
    \includegraphics[width=.6\textwidth ]{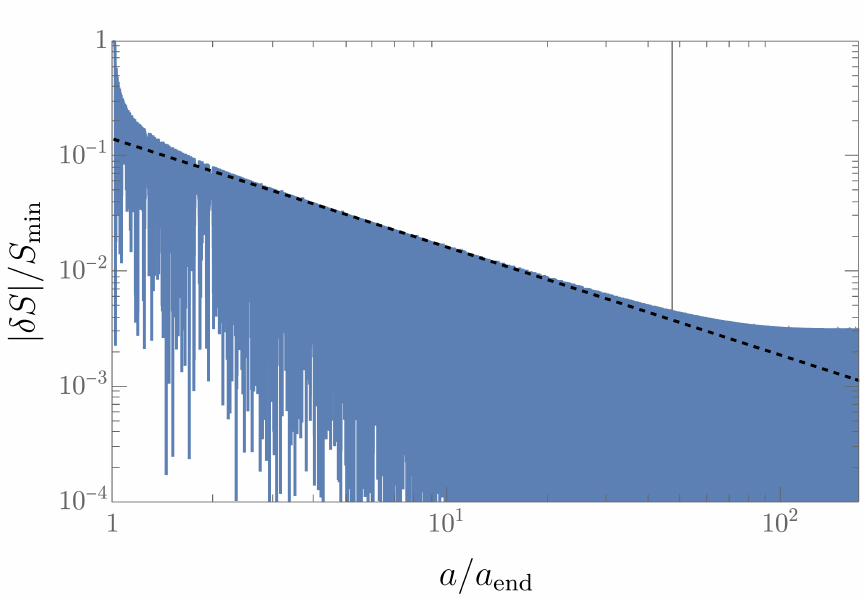}
    \caption{%
        Time evolution of $|\delta S|/S_\mathrm{min}$ for $\mathcal{H}_\mathrm{end} = 10^9$\,GeV, $T_\mathrm{R} = 10^{12}$\,GeV, $\mu_S = 10^5$\,GeV, $c = 1$, $\lambda_S = 0.2$, and $\lambda_{SH} = 0.5$.
        The horizontal axis and the vertical gray line are the same as in Fig.~\ref{fig: time evolution of s}.
        The black-dashed line is proportional to $a^{-15/16}$ (see \eqref{eq: deltaS/Smin}).
    }
    \label{fig: time evolution of dS/Smin}
\end{figure}

We show the numerical result in Fig.~\ref{fig: time evolution of s}.
Here, we set $\mathcal{H}_\mathrm{end} = 10^9$\,GeV, $T_\mathrm{R} = 10^{12}$\,GeV, $\mu_S = 10^5$\,GeV, $c = 1$, $\lambda_S = 0.2$, and $\lambda_{SH} = 0.5$.
After inflation, $s_\mathrm{min}$ (black-dashed line) rapidly increases, and $s$ (blue line) follows $s_\mathrm{min}$ with a slight delay.
Then, $s$ oscillates around $s_\mathrm{min}$ with a decreasing amplitude and converges to $s_\mathrm{min}$.
To see the convergence to $s_\mathrm{min}$, we also show $|\delta S|/S_\mathrm{min} = |s - s_\mathrm{min}|/s_\mathrm{min}$ in Fig.~\ref{fig: time evolution of dS/Smin}.
While the amplitude of $|\delta S|/S_\mathrm{min}$ rapidly decreases soon after inflation, its evolution is well represented by \eqref{eq: deltaS/Smin} after that.
We expect that the decrease in $|\delta S|/S_\mathrm{min}$ during the reheating epoch is a general outcome since its derivation does not depend on a specific assumption on the model parameters.
We show some results for other parameters in App.~\ref{app:numerical S}, which further support our argument.
Consequently, we have confirmed that $S$ does not overshoot the origin after inflation.
In contrast, if we neglect the thermal potential, $S$ overshoots the origin although $S_\mathrm{min}$ is always nonzero.

Note that we have neglected the thermal dissipation effect on the zero mode of $S$. Including the dissipation effect would reduce the oscillation amplitude, which makes our conclusion that $S$ does not overshoot the origin more robust.
On the other hand, $S$ may be thermalized around its temporal potential minimum. The magnitude of thermal fluctuations is about $\sim T$, which can be always smaller than $S_{\rm min}$ for $\lambda_S \lesssim \lambda_{SH}$.\footnote{
    Even if $T \sim S_{\rm min}$, thermal fluctuation does not lead to domain wall problem due to the population bias~\cite{Lalak:1994qt,Larsson:1996sp,Gonzalez:2022mcx,Kitajima:2023kzu}.
}

Therefore, the CP symmetry continues to be broken during and after inflation and there is no symmetry restoration after inflation. There is no domain wall problem in this model.

\subsection{Leptogenesis}

Now we discuss the leptogenesis~\cite{Fukugita:1986hr,Giudice:2003jh,Buchmuller:2004nz} in our scenario. 
In the BBP model, we can explain the origin of the CP violation in the quark and lepton sectors simultaneously by adopting a $Z_4$ symmetry~\cite{Branco:2003rt}.
Here, we show that this framework is compatible with our model with the $Z_4$ and $Z_{4n}^{\rm (app)}$ symmetries.
The lepton sector of the model is given by
\begin{align}
	-\mathcal L = \frac{1}{2}M^N_{ij} \overline{N^c_i} N_j  + y_{ij}^\nu \widetilde H \overline L_i N_j + y_{ij}^e H \overline L_i e_{Rj} + {\rm h.c.},
\end{align}
where $N_i, L_i, e_{Ri}$ are right-handed neutrinos, the Standard Model lepton doublets, right-handed leptons respectively, and
\begin{align}
	M^N_{ij} = \xi_{ij}S + \xi'_{ij} S^*,
\end{align}
with $\xi_{ij}, \xi_{ij}', y_{ij}^\nu, y_{ij}^e$ being real constants.
Charge assignments are listed in Table~\ref{table}. All the terms in the lepton sector are consistent with the $Z_4$ and $Z_{4n}^{\rm (app)}$ symmetries.
After having VEV of $S$ as $S=v_S e^{i\theta}$, the right-handed neutrino mass matrix $M^N_{ij}$ becomes complex, while Yukawa couplings remain real. 
After diagonalizing the right-handed neutrino mass matrix $M_{ij}^{N}$ and making its eigenvalues real through the unitary transformation as $N_i \to U^{(N)}_{ij} N_j$, the Yukawa couplings become complex: $\widetilde y^\nu_{ij} = y^\nu_{ik} U^{(N)\dagger}_{kj}$.
Thus there are enough number of CP phases in the lepton sector for producing the matter-antimatter asymmetry.
The light neutrino mass generation through the seesaw mechanism also works~\cite{Minkowski:1977sc,Yanagida:1979as,Gell-Mann:1979vob}.

Assuming one of the right-handed neutrinos ($N_1$) is much lighter than the others, the asymmetry parameter is given by~\cite{Hamaguchi:2002vc}
\begin{align}
	\frac{\Gamma(N_1\to L + H) - \Gamma(N_1\to \overline L + \overline H) }{\Gamma(N_1\to L + H) + \Gamma(N_1\to \overline L + \overline H)}
	&\simeq -\frac{3}{16\pi (\tilde y^\nu\tilde y^{\nu\dagger})_{11}}\sum_{i=2,3}{\rm Im}\left[ (\tilde y^\nu\tilde y^{\nu\dagger})^2_{1i} \right] \frac{M_1}{M_i}\\
	&\simeq 1\times 10^{-6}\left(\frac{M_1}{10^{10}\,{\rm GeV}}\right)\left(\frac{m_{\nu 3}}{0.05\,{\rm eV}}\right)\delta,
\end{align}
where $M_i$ is the mass of $N_i$, $m_{\nu 3}$ is the heaviest active neutrino mass, and $\delta$ is the effective CP phase parameter, which is $\mathcal O(1)$ in the present model. 
Note that in the scenario of Sec.~\ref{sec:DW}, $S$ (and hence $M_1$ and $m_{\nu 3}$) is time dependent.
For simplicity, we assume $M_1 \lesssim v_S$ so that the $S$ field is fixed to its zero temperature value when $N_1$ is decoupled from thermal bath and decays.
Then the final baryon asymmetry is given by
\begin{align}
	\frac{n_B}{s} \simeq 2\times 10^{-10}\,\kappa\left(\frac{M_1}{10^9\,{\rm GeV}}\right)\left(\frac{m_{\nu 3}}{0.05\,{\rm eV}}\right)\delta,
\end{align}
where $\kappa$ represents the suppression factor, which is at most $\mathcal O(0.1)$, due to either strong washout effect or smallness of the right-handed neutrino abundance~\cite{Buchmuller:2004nz}.
It is maximized if $N_1$ is thermalized and decays around when it decouples from thermal plasma at $T\sim M_1$. 
We can obtain the correct baryon asymmetry if $M_1 \gtrsim 10^9$\,GeV and $T_{\rm R} \gtrsim 10^9$\,GeV.
As we have already seen, the VEV of $S$ much higher than $10^9$\,GeV is consistent with the ``quality'' of the Nelson-Barr mechanism in our model, and also the temperature much higher than $10^9$\,GeV does not always lead to domain wall formation.

Finally, we comment on other models for leptogenesis in the framework of the Nelson-Barr mechanism.
Ref.~\cite{Asadi:2022vys} considered ``chiral Nelson-Barr model'' to solve these problems, in which chiral U(1) symmetry is newly introduced to suppress dangerous operators contributing to $\theta_s$. To avoid the anomaly, additional chiral fermions and scalar have been introduced.
Ref.~\cite{Suematsu:2023jqa} realizes leptogenesis with a relatively low reheating temperature by extending the discrete symmetry and introducing additional vector-like leptons and scalars to the BBP model.

\section{Conclusions and discussion}
\label{sec:conc}

In this paper, we revisited the minimal Nelson-Barr model by Bento, Branco, and Parada~\cite{Bento:1991ez} and proposed a variation of the BBP model avoiding the quality problem.
We impose an additional approximate symmetry without introducing any additional field contents.
Due to the additional symmetry, the mass term and Yukawa coupling of the vector-like quarks are suppressed, and the dangerous contributions to the $\theta$ parameter from higher-dimensional operators and one-loop effects are also suppressed.
Consequently, the strong CP problem can be solved even with a high scale of the spontaneous CP breaking $\gg 10^8$\,GeV, in contrast to the original BBP model.

Moreover, we discussed some cosmological implications.
To avoid the domain wall problem due to post-inflationary CP breaking, we propose the possibility of a negative thermal potential for the CP-violating scalar.
This ensures that the CP symmetry is always spontaneously broken during and after inflation even if the reheating temperature is much higher than the CP breaking scale.
We also explored leptogenesis in this framework.
By introducing right-handed neutrinos and imposing nontrivial charges under the discrete symmetries to the lepton sector, the right-handed neutrinos are coupled to the Standard Model Higgs and the CP-violating scalar through the Yukawa terms.
Then, the right-handed neutrino decays in a CP-violating way, and the thermal leptogenesis works.
Here, successful leptogenesis requires a high reheating temperature, $T_\mathrm{R} \gtrsim 10^9$\,GeV, which can be consistently realized in our model.
Simultaneously, the light neutrino masses can be explained through the seesaw mechanism.
The compatibility with a high reheating temperature also indicates the possibility of high-scale inflation. This is in contrast to the original BBP model, in which the inflationary scale and the reheating temperature are limited as $\lesssim 10^8$\,GeV.
In addition, the mixing of the Standard Model quarks and vector-like quarks can be probed via the non-unitarity of the CKM matrix.

Finally we mention possible implications for dark matter. Without any further assumptions, there is no dark matter candidate in our model. Of course, one can add dark matter sector by hand. However, here we mention a possibility of dark matter without extending the particle content of our model. 

One possibility is to regard one of the right-handed neutrinos as dark matter. To do so, we impose an additional $Z_2$ symmetry under which only $N_3$ is odd and all other fields are even. Then the Yukawa coupling of $N_3$ is forbidden and it is stable so that it is a good dark matter candidate~\cite{Okada:2010wd}.
The seesaw mechanism still works with two right-handed neutrinos $N_1, N_2$ and the leptogenesis scenario also works~\cite{Frampton:2002qc,Ibarra:2003up}.
A prediction of this scenario is that the lightest active neutrino is massless, which can in principle be tested by the neutrinoless double beta decay experiment and the cosmological measurement of the absolute neutrino masses.
In our model, $N_3$ can be produced in either two ways: production through the coupling with $S$ or the gravitational production. The coupling with $S$ can be forbidden by modifying the $Z_4$ charge of $N_3$.
Then only the universal production process is the gravitational one. The gravitational production of fermions has been studied in Refs.~\cite{Chung:2011ck,Ema:2015dka,Ema:2019yrd,Clery:2021bwz}.
The dark matter abundance is then given by~\cite{Ema:2019yrd}
\begin{align}
    \Omega_{\rm DM} h^2 \simeq \begin{cases}
    \displaystyle 0.1\,\left(\frac{M_{3}}{10^{10}\,{\rm GeV}}\right)\left(\frac{T_{\rm R}}{10^{11}\,{\rm GeV}}\right)\left(\frac{\mathcal H_{\rm end}}{10^{10}\,{\rm GeV}}\right)\left(\frac{M_{3}}{m_{\phi}}\right)^2
    & {\rm for}~~M_{3} > \mathcal H_{\rm end}\\
    \displaystyle 0.1\,\left(\frac{M_{3}}{10^{10}\,{\rm GeV}}\right)^2\left(\frac{T_{\rm R}}{10^{11}\,{\rm GeV}}\right) & {\rm for}~~M_{3} < \mathcal H_{\rm end}
    \end{cases},
\end{align}
where $m_\phi$ is the inflaton mass. Thus it can be consistent with the observed dark matter abundance for reasonable parameter choice.

Another possibility is to make $\sigma$, the phase component of $S$, very light like the axion so that it can be a dark matter candidate~\cite{Dine:2024bxv}.
For realizing this scenario, we need to impose a global U(1) symmetry so that the phase of $S$ is a pseudo Nambu-Goldstone boson. One should note that either $g_i$ or $g_i'$ must be zero for each $i$ to be consistent with the U(1) symmetry. In order to keep the effective CP phase, we need to impose a flavor-dependent U(1) charge on quarks which leads to a special structure like, e.g., $g_i\propto (1,0,0)$ and $g'_i\propto (0,1,0)$.
However, additional ingredients are required for obtaining the realistic CKM matrix and so on~\cite{Dine:2024bxv}. Thus we do not discuss this possibility further.

\section*{Acknowledgment}

This work was supported by World Premier International Research Center Initiative (WPI), MEXT, Japan.
This work was also supported by JSPS KAKENHI (Grant Numbers 24K07010 [KN], 23KJ0088 [KM], and 24K17039 [KM]).

\appendix

\section{One loop correction to the strong CP angle}
\label{app:oneloop}

We derive one-loop correction to the strong CP angle, following Ref.~\cite{Bento:1991ez}.

We define $d_{\alpha}$ with $\alpha=1$--$4$ such that $d_{i}$ $(i=1$--$3)$ denotes the Standard Model down quarks and $d_{4} = D$ denotes the additional heavy quark.
The tree level mass matrix (\ref{massmatrix}) can be diagonalized with the bi-unitary transformation, $d_{R\alpha} = U_{\alpha\beta} d'_{R\beta}$ and  $d_{L\alpha} = V_{\alpha\beta} d'_{L\beta}$, where the prime indicates the (tree-level) mass eigenstates.
The mass matrix in this basis is given by
\begin{align}
	&\mathcal M' = \mathcal M'_0 + \mathcal M'_1,\\
	&\mathcal M'_0 =V^\dagger \mathcal M U = {\rm diag}(m_d,m_s, m_b, m_D),
\end{align}
where $m_d,m_s, m_b, m_D$ are down, strange, bottom, and additional heavy quark masses, respectively, and $\mathcal M_1'$ denotes the one-loop correction. 
Note that the tree-level mass eigenvalues $m_\alpha$ are made real and positive without introducing the strong CP angle, since ${\rm det}\mathcal M$ is real.  
$\mathcal M_1'$ may contain complex phases and lead to the strong CP angle.
In general, $\mathcal M_1'$
has both diagonal and off-diagonal components, but the dominant contribution to $\bar\theta_s={\rm arg}[{\rm det}\mathcal M]$ comes from diagonal components of $\mathcal M_1'$, since at least two entries are required for the off-diagonal components to contribute to the determinant, while we need only one component for the diagonal component.
Thus below we focus on diagonal components of $\mathcal M_1'$.
In this case, the correction to the mass matrix is simply given by $m_\alpha \to m_\alpha + \delta m_\alpha$ and the strong CP phase is given by
\begin{align}
	\bar\theta_s \simeq {\rm Im} \left[\sum_{\alpha} \frac{\delta m_{\alpha}}{m_\alpha}\right] = \sum_{\alpha} \frac{{\rm Im}(\delta m_{\alpha})}{m_\alpha},
\end{align}
where we have defined the one-loop correction to the diagonal component as
\begin{align}
	 -\mathcal L = \overline d'_{L\alpha} (m_\alpha + \delta m_\alpha) d'_{R\alpha} + {\rm h.c.} 
	 =  \overline d'_\alpha (m_\alpha + \Sigma_\alpha) d'_\alpha,
\end{align}
where
\begin{align}
	\Sigma_\alpha =  {\rm Re}(\delta m_\alpha) + i\gamma_5  {\rm Im}(\delta m_\alpha).
\end{align}
\begin{figure} 
\begin{center}
\begin{tikzpicture}\begin{feynman}
\vertex (d1) {\(d'_{\alpha}\)}; \vertex[right=2cm of d1] (d2); \vertex[right=1.5cm of d2] (d3); \vertex[right=1.5cm of d3] (d4); \vertex[right=2cm of d4] (d5) {\(d'_{\alpha}\)};
\diagram* { 
(d1)--[momentum=\(p\)](d2),(d2)--[edge label' =\(d'_{\beta}\),momentum=\(p-k\)] (d4),(d4)--[momentum=\(p\)](d5),
(d2)--[scalar, half right, momentum'=\(k\), edge label=\(\phi'_A\)](d4)
};
\end{feynman} \end{tikzpicture}
\end{center}
\caption{One loop correction to the self energy diagram for fermions $d_{\alpha}$ with $\alpha=1$--$4$ including the Standard Model down quarks $(\alpha=1$--$3)$ and an additional heavy quark $(\alpha=4)$. They are mass eigenstates at the tree level.}
\label{fig:1loop}
\end{figure}

Fig.~\ref{fig:1loop} shows a diagram that contributes to ${\rm Im}(\delta m_\alpha)$.
To evaluate this, we write Yukawa couplings in the mass basis. By expanding $S= (v_S + (s+i\sigma)/{\sqrt 2})e^{i\theta} $ and $H =(0, v_H + h/\sqrt{2})$, the Yukawa couplings are written as
\begin{align}
    -\mathcal L 
    = \sum_{a,A=1}^3 \phi'_A R^A_a \left[ \overline d'_{L\alpha}\Gamma'^a_{\alpha\beta} d'_{R\beta} + {\rm h.c.} \right]
    =  \sum_{a,A=1}^3 \phi'_A R^A_a \left[ \overline d'_{\alpha}\widetilde\Gamma'^a_{\alpha\beta} d'_{\beta} \right],
\end{align}
where we defined
\begin{align}
	\phi_a=\begin{pmatrix}h \\ s \\ \sigma \end{pmatrix},~~~\phi'_A=R_A^a \phi_a,
\end{align}
with prime denoting the mass eigenstate and $R_A^a$ being a real orthogonal matrix, and 
\begin{align}
	&\Gamma^h_{\alpha\beta}=\frac{1}{\sqrt 2}\begin{pmatrix}y_{ij} & 0 \\ 0 & 0 \end{pmatrix},~~~
	\Gamma^s_{\alpha\beta}=\frac{1}{\sqrt 2}\begin{pmatrix} 0 & 0 \\ g_ie^{i\theta} + g'_ie^{-i\theta}  & 0 \end{pmatrix},~~~
	\Gamma^\sigma_{\alpha\beta}=\frac{1}{\sqrt 2}\begin{pmatrix} 0 & 0 \\ i(g_ie^{i\theta} - g'_ie^{-i\theta})  & 0 \end{pmatrix}, \nonumber \\
	&\Gamma'^a_{\alpha\beta} = V^\dagger_{\alpha\gamma}\Gamma^a_{\gamma\delta}U_{\delta\beta},
	\widetilde\Gamma'^a_{\alpha\beta} =\Gamma'^{a{\rm (H)}}_{\alpha\beta} + \gamma_5 \Gamma'^{a{\rm (AH)}}_{\alpha\beta},
\end{align}
where $X^{\rm (H)} \equiv (X+X^\dagger)/2$ and $X^{\rm (AH)} \equiv (X-X^\dagger)/2$ for general matrix $X$.

The one-loop self-energy is calculated as
\begin{align}
	\Sigma_\alpha = - i\sum_{A,\beta} R_A^a R_A^b\, \widetilde\Gamma'^a_{\alpha\beta}\, I_{A,\alpha,\beta}(\slashed{p})\, \widetilde\Gamma'^b_{\beta \alpha},
\end{align}
where
\begin{align}
	I_{A,\alpha,\beta}(\slashed{p}) 
	&= \int\frac{d^4k}{(2\pi)^4} \frac{i(\slashed{p}-\slashed{k}+m_\beta)}{(p-k)^2-m_\beta^2} \frac{i}{k^2-m_A^2}\\
	&=-\int\frac{d^4 \ell}{(2\pi)^4}\int_0^1 dxdy\delta(x+y-1) \frac{\slashed{\ell}+(1-x)\slashed{p}+m_\beta}{\left[\ell^2 - (x^2p^2+ y m_A^2-x(p^2-m_\beta^2))\right]^2} \\
	&= \frac{i}{16\pi^2}\int_0^1 dx\left[(1-x)\slashed{p} + m_\beta \right]\log\left[\frac{x^2 m_\alpha^2 +(1-x)m_A^2 -x(m_\alpha^2-m_\beta^2)}{\Lambda^2}\right],
\end{align}
where we have defined $\ell \equiv k + px$ in the second line and used $p^2=m_\alpha^2$ in the last line, and $\Lambda$ is the renormalization scale. 
By noting $\slashed{p}\widetilde\Gamma' = (\widetilde\Gamma')^\dagger \slashed{p}$, the term proportional to $\slashed{p}$ is Hermitian and does not contribute to ${\rm Im}(\delta m_\alpha)$.
Thus
\begin{align}
	&\sum_\alpha \frac{\Sigma_\alpha}{m_\alpha} = \sum_{A,\alpha,\beta}\frac{1}{16\pi^2} R_A^a R_A^b  m_\alpha^{-1}\left[\widetilde\Gamma'^a_{\alpha\beta} m_\beta \widetilde\Gamma'^b_{\beta\alpha}\right] J_{A,\alpha,\beta}, \\
	&J_{A,\alpha,\beta} = \int_0^1 dx\log\left[\frac{x^2 m_\alpha^2 +(1-x)m_A^2 -x(m_\alpha^2-m_\beta^2)}{\Lambda^2}\right].
\end{align}
Picking up only $\gamma_5$-dependent terms, we obtain
\begin{align}
	\bar\theta_s = \sum_{A,\alpha,\beta}\frac{1}{16\pi^2} R_A^a R_A^b  m_\alpha^{-1}
 \mathrm{Im}
 \left[\widetilde\Gamma'^{a{\rm (H)}}_{\alpha\beta} m_\beta \widetilde\Gamma'^{b{\rm (AH)}}_{\beta\alpha}
 + \widetilde\Gamma'^{a{\rm (AH)}}_{\alpha\beta} m_\beta \widetilde\Gamma'^{b{\rm (H)}}_{\beta\alpha}\right] J_{A,\alpha,\beta}.
\end{align}

This expression is significantly simplified by approximating $J_{A,\alpha,\beta}$ as
\begin{align}
	J_{A,\alpha,\beta} \simeq \int_0^1 dx \log\left(\frac{m_A^2}{\Lambda^2}\right)=\log\left(\frac{m_A^2}{\Lambda^2}\right),
\end{align}
for $m_A \gg m_\alpha, m_\beta$. Using $m_\alpha = U^\dagger \mathcal M V = V^\dagger \mathcal M^\dagger U$, we obtain
\begin{align}
	\bar\theta_s =\frac{1}{16\pi^2}\sum_A R_A^a R_A^b {\rm Im }\left[ {\rm Tr}\left[ \mathcal M^{-1} \Gamma^a \mathcal M^{\dagger} \Gamma^{b}\right] \right]\log\left(\frac{m_A^2}{\Lambda^2}\right).
\end{align}
By substituting the concrete forms of $\Gamma^a$ and $\mathcal M$, we find that only the combination $(a,b)=(h,\sigma)$ contributes to $\theta_s$.\footnote{
    Note that $\mathcal M^{-1}$ is given by
    \begin{align} \mathcal M^{-1} = \begin{pmatrix} m_{ij}^{-1} & 0 \\ -B_i m^{-1}_{ij}M^{-1} & M^{-1} \end{pmatrix} \end{align} 
} The result is
\begin{align}
	 \bar\theta_s = \frac{1}{32\pi^2}\frac{v_S}{v_H}\sum_{i,A}(g_i^2 - g_i'^2) R_A^h R_A^\sigma \log\left(\frac{m_A^2}{\Lambda^2}\right).
\end{align}
The factor $R_A^h R_A^\sigma$ roughly corresponds to the mixing angle between $h$ and $\sigma$, which should be of the order of $\sim \gamma_{SH} \sin(2\theta)v_H/v_S$, where $\gamma_{SH}$ is the four-point coupling $V = \gamma_{SH}|H|^2(S^2+S^{*2})$.
Noting also that $\sum_A R_A^h R_A^\sigma = 0$ from the orthogonal property of the matrix $R_A^a$, we find
\begin{align}
	 \bar\theta_s 
  \sim 
  \frac{1}{32\pi^2}\gamma_{SH}\sin(2\theta)\sum_i(g_i^2 - g_i'^2)\log\left(\frac{m_h^2}{m_\sigma^2}\right).
\end{align}

\section{Numerical results for \texorpdfstring{$\delta S$}{}}
\label{app:numerical S}

\begin{figure}[t]
    \centering
    \includegraphics[width=.45\textwidth ]{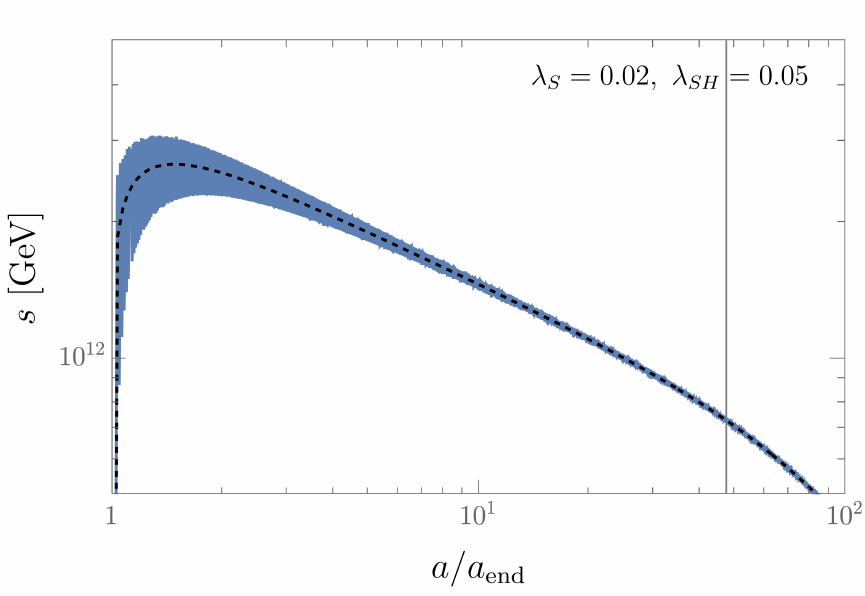}
    \hspace{5mm}
    \includegraphics[width=.45\textwidth ]{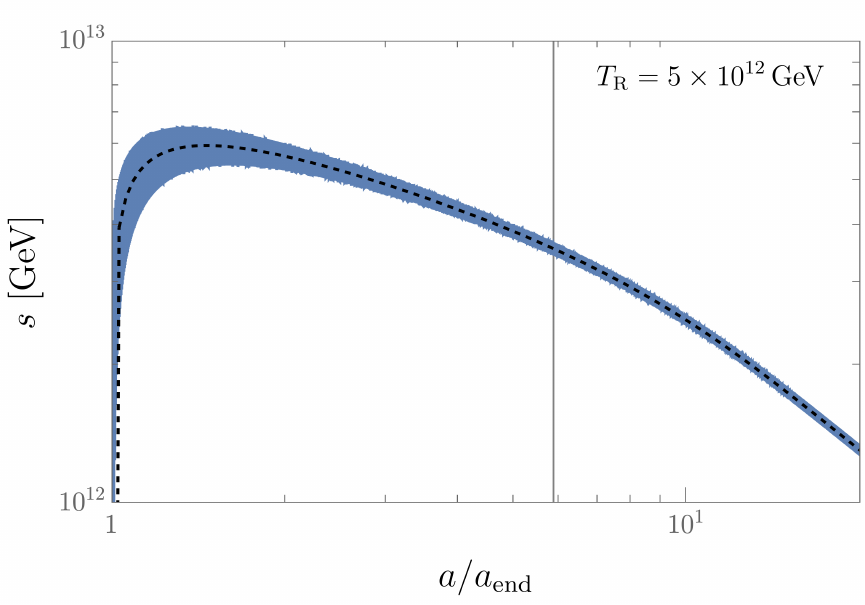}
    \caption{%
        Time evolution of $s$ for the same parameters as in Fig.~\ref{fig: time evolution of s} except for the parameters given in the plots.
        The vertical gray line corresponds to $T = T_\mathrm{R}$, and the black-dashed lines show $s_\mathrm{min}$.
    }
    \label{fig: time evolution of s in App}
\end{figure}

Here, we provide the numerical results for the time evolution of the scalar field, $S$, for other parameter sets than that in the main text.
In Fig.~\ref{fig: time evolution of s in App}, we show the time evolution of $s$ for smaller $\lambda_S$ and $\lambda_{SH}$ (left panel) and larger $T_\mathrm{RH}$ (right panel) than in Fig.~\ref{fig: time evolution of s}.
In both cases, $s$ approaches $s_\mathrm{min}$ with damped oscillations.
We have also checked that the oscillation amplitude of $|\delta S|/S_\mathrm{min}$ decreases as $\propto a^{-15/16}$ during reheating.

\begin{figure}[t]
    \centering
    \includegraphics[width=.45\textwidth ]{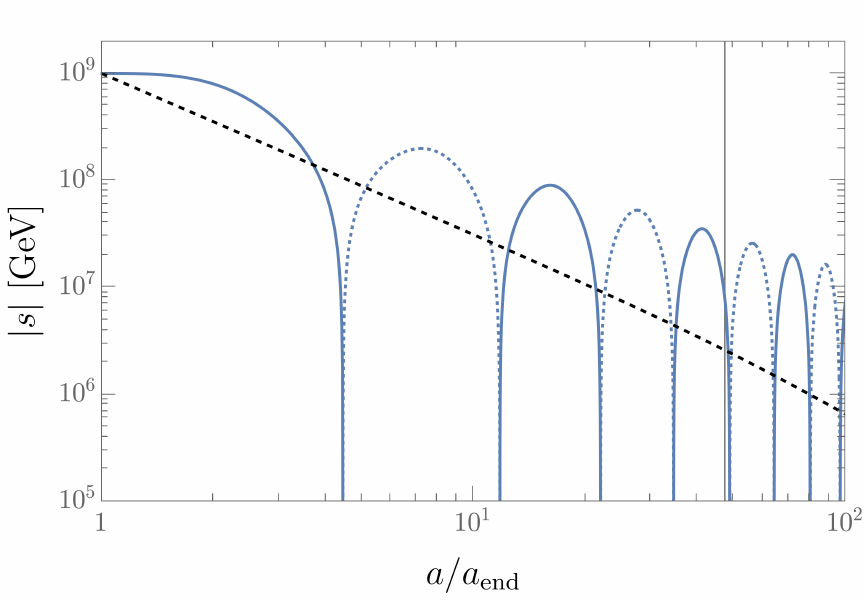}
    \caption{%
        Time evolution of $s$ for the same parameters as in Fig.~\ref{fig: time evolution of s} except for $\lambda_{SH} = 0$.
        The dotted line represents a negative sign of $s$.
        The vertical gray line corresponds to $T = T_\mathrm{R}$, and the black-dashed lines show $s_\mathrm{min}$.
    }
    \label{fig: time evolution of s overshoot}
\end{figure}

In Fig.~\ref{fig: time evolution of s overshoot}, we show the time evolution of $s$ without the thermal potential, i.e., $\lambda_{SH} = 0$.
In contrast to the above cases, $s$ overshoots the origin and oscillates around $s = 0$.

\bibliographystyle{utphys}
\bibliography{ref}

\end{document}